\documentclass[a4paper]{llncs}
\usepackage{makeidx}  % allows for indexgeneration
\usepackage{lmodern}% http://ctan.org/pkg/lm
\usepackage[T1]{fontenc}
\usepackage{amsmath}

\usepackage{proving}

\begin{document}

\title{
   Extending E Prover with Similarity Based Clause Selection Strategies
}

\titlerunning{
   Extending E Prover with Similarity Based Clause Selection Strategies
}

\author{
    Jan Jakub\r{u}v\thanks{Supported by the ERC Consolidator grant nr. 649043 \textit{AI4REASON}.}
\and
    Josef Urban${}^\star$
}
\institute{
    Czech Technical University,
    CIIRC, Prague\\
   \email{\{jakubuv,josef.urban\}@gmail.com}
}
\authorrunning{Jakub\r{u}v, Urban}

\maketitle

\begin{abstract}

E prover is a state-of-the-art theorem prover for first-order logic
with equality.  E prover is built around a saturation loop, where new
clauses are derived by inference rules from previously derived
clauses.  Selection of clauses for the inference provides the main
source of non-determinism and an important choice-point of the loop
where the right choice can dramatically influence the proof search. In
this work we extend E Prover with several new clause selection
strategies based on similarity of a clause with the conjecture.  In
particular, clauses which are more related to the conjecture are
preferred. We implement different strategies that define the
relationship with a conjecture in different ways.  We provide an
implementation of the proposed selection strategies and we evaluate
their efficiency on an extensive benchmark set.

\keywords{Automated Theorem Proving, Large Theory Reasoning, Clause Selection}
\end{abstract}

\setlength{\tabcolsep}{.4em}

\section{Introduction}

Many state-of-the-art automated theorem provers (ATPs) are based on the \emph{given clause algorithm}
introduced by \emph{Otter}~\cite{mccune1994otter}.
The input problem $T\cup\{\lnot C\}$ is translated into a refutationally equivalent
set of clauses.
Then the search for a contradiction, represented by the empty clause, is
performed maintaining two sets: the set $P$ of \emph{processed clauses} and the set
$U$ of \emph{unprocessed} clauses.
Initially, all the input clauses are unprocessed.
The algorithm repeatedly selects a \emph{given clause} $g$ from $U$ and
generates all possible inferences using $g$ and the processed clauses from $P$.
Then, $g$ is moved to $P$, and $U$ is extended with the newly produced
clauses.
This process continues until a resource limit is reached, or the empty clause is inferred, or $P$ becomes
\emph{saturated}, that is, nothing new can be inferred.

The search space of this loop grows quickly.
Several methods can be used to make the proof search more efficient.
The search space can be narrowed by adjusting (typically restricting) the inference rules, pruned by
using \emph{forward} and \emph{backward subsumption}, reduced by
pre-selecting relevant input clauses, or otherwise simplified.
One of the main sources of non-determinism affecting efficiency of the
search is the selection of the given clause.
Clever selection mechanism can improve the search dramatically: in
principle, one only needs to do the inferences that participate in the
final proof. So far, this is often only a tiny portion of all the
inferences done by the ATPs during the proof search.

\section{Clause Selection in E Prover}
\label{sec:anl}

E~\cite{schulz2002brainiac} is a state-of-the-art theorem prover which we
use as a basis for implementation.
%%%E implements \emph{clauses} as sets of equations.
%%%An \emph{equation} is either a positive $\termA\simeq\termB$, or a negative
%%%$\termA\not\simeq\termB$, pair of terms.
%%%A classical first-order predicate $p(\termA)$ (or its negation $\lnot
%%%p(\termA)$) is represented using a special $\true$ constant as
%%%$p(\termA)\simeq\true$ (or $p(\termA)\not\simeq\true$).
%%%Hence E does not distinguish between terms and predicates, however,
%%%this information is internally preserved as a term flag.
%
The selection of a given clause in E is implemented by a combination of
priority and weight functions.
A \emph{priority function} assigns an integer to a clause and is used to
pre-order clauses for weight evaluation.
A \emph{weight function} takes additional specific arguments
and assigns to each clause a real number called \emph{weight}.
A \emph{clause evaluation function} $\CEF$ is specified
by a priority function, weight function, and its arguments.
Each $\CEF$ selects the clause with the smallest pair
$(\textit{priority}, \textit{weight})$ for inferences.
%
%Each E heuristic $\textit{\procStyle{H}}$ is hence specified 
%using the syntax
%``\heur{\textit{WeightFunction}}{\textit{PriorityFunction,\ldots}}''
%with a variable number of comma separated arguments of the weight function.
E allows a user to select an \emph{expert heuristic} on a command line
in the format
   ``\texttt{($n_1$*$\CEF_1$,\ldots,
      $n_k$*$\CEF_k$)}'',
where integer $n_i$ indicates how often the corresponding 
$\CEF_i$ should be
used to select a given clause. 
E additionally supports an \emph{autoschedule} mode where several expert
heuristics are tried, each for a selected time period.
The heuristics and time periods are automatically chosen based on 
input problem properties.

One of the well-performing weight functions in E, which we also use as a reference for
evaluation of our weight functions, is the \emph{conjecture symbol weight}.
This weight function counts symbol occurrences with different weights based
on their appearance in the conjecture as follows.
Different weights $\COEFfunc$, $\COEFconst$, $\COEFpred$,
and $\COEFvar$ are assigned % in turn 
to function, constant,
and predicate symbols, and to variables.
The weight of a symbol which appears in the conjecture is multiplied by
$\COEFconj$, typically $\COEFconj<1$ to prefer clauses with conjecture
symbols.
To compute a term weight, the given symbol weights are summed for all symbol
occurrences.
This evaluation is extended to equations and to clauses.
%, by summing weights of both sides,
%and to clauses, by summing weights of the equations.
%Arguments $\COEFmaxterm$, $\COEFmaxlit$, $\COEFposmult$ specify additional
%multipliers applied % in turn 
%to a maximal term in a literal (w.r.t. a
%selected ordering), to the maximal literal in a clause, and to a positive
%equality.

%%
%%
%%\heurcodeSim{ConjectureRelativeSymbolWeight}{$\COEFconj$,$\COEFfunc$,$\COEFconst$,$\COEFpred$,$\COEFvar$,}
%%
%%
%%\[
%%   \weight{\clause} = \sum_{\lit\in\clause}\delta_\lit*\weight{\lit}
%%   \quad\mbox{where}\quad
%%   \delta_\lit = 
%%   \begin{cases}
%%      \COEFmaxlit & \mbox{iff $\lit$ is maximal in $\clause$ }\\
%%      1 & \mbox{otherwise}
%%   \end{cases}
%%\]
%%
%%\small
%%\[
%%   \weight{\termA\not\simeq\termB} = 
%%      \COEFmaxterm*\weight{\termA}+\delta_\termB*\weight{\termB}
%%   \quad\mbox{where }
%%   \delta_\termB = 
%%   \begin{cases}
%%      1 & \mbox{iff $\termB\leq\termA$ }\\
%%      \COEFmaxterm & \mbox{otherwise}
%%   \end{cases}
%%\]
%%\normalsize
%%
%%\[
%%   \weight{\termA\simeq\termB} = \COEFposmult*\weight{\termA\not\simeq\termB}
%%\]
%%

\section{Similarity Based Clause Selection Strategies}
\label{sec:weights}

Many of the best-performing weight functions in E are based on a similarity of a
clause with the conjecture, for example, the \emph{conjecture symbol weight}
from the previous section.
In this paper we try to answer the question whether or not it makes sense to
also investigate a term structure.
We propose, implement, and evaluate several weight functions
which utilize conjecture similarity in different ways. Typically they extend
the symbol-based similarity by similarity on terms. Using finer formula
features improves the high-level premise selection
task~\cite{kaliszyk2015efficient}, which motivates  this work on steering
also the internal selection in E.
%Our heuristics definitions share the syntax
%``$\heurcode{\textit{WeightName}}{\ldots,}$'' where the last three arguments are
%used as in the previous section and the others are as follows.
We first describe the common arguments of our weight functions and then
function-specific properties.

\subsubsection{Common Arguments ($\mathsf{\mathit{v}}$,$\mathsf{\mathit{r}}$,$\mathsf{\mathit{e}}$)}

We implement two ways of term variable
normalization, selected by the argument $\vartt{v}$. 
Either 
(1) variables are $\alpha$-normalized, naming them consistently by
their appearance in the term from left to right (value ``$\Alf$''), or 
(2) all variables are unified to a single variable (``$\Uni$''). 
This provides differently coarse notions of similarity.
Each of our weight functions relates a term to the global set $\related$.
This set $\related$, controlled by the argument $\vartt{r}$, contains either 
(1) all conjecture terms (``$\Ter$''),
(2) conjecture terms and their subterms (``$\Sub$''),
(3) conjecture subterms and top-level generalizations (``$\Top$''), or to
(4) conjecture subterms and all their generalizations (``$\Gen$'').
Each of our weight functions implements a different function $\WEIGHTONE$ which assigns
a weight to a term.
We use three different ways of extending $\WEIGHTONE$ to compute a term
weight, selected by the argument $\vartt{e}$.
Either 
(1) $\WEIGHTONE$ value is used directly (value ``$\Sim$''), or 
(2) values of $\WEIGHTONE$ for all the subterms are summed (``$\Sum$''), or 
(3) the maximal value of $\WEIGHTONE$ on all of the subterms is used
(``$\Max$''). 

%The next subsections describe the new heuristics and their evaluation is
%provided in the next section.

%%%Let $\subterms{\termA}$ denote the set of all subterms of $\termA$
%%%(including $\termA$).
%%%
%%%%\small
%%%\[\begin{array}{lcl}
%%%   \weightsum{\termA} = \sum_{\termB\in\subterms{\termA}}{\weightone{\termB}}
%%%   &\quad&
%%%   \weightmax{\termA} = \max_{\termB\in\subterms{\termA}}{\weightone{\termB}}
%%%\end{array}\]
%%%\normalsize

\subsubsection{Conjecture Subterm Weight (\Term)}

The first of our weight functions  is similar to the standard \emph{conjecture symbol
weight}, counting instead of symbols the number of subterms a term shares with the conjecture.
The weight function \Term takes
five specific arguments $\COEFconj$, $\COEFfunc$, $\COEFconst$, $\COEFpred$
and $\COEFvar$ and $\weightbase{\Term}$ equals weight $\COEFfunc$ for functional terms,
$\COEFconst$ for constants, $\COEFpred$ for predicates, and $\COEFvar$ for
variables, possibly multiplied by $\COEFconj$ when $\termA\in\related$.

%%\[
%%   \weightzero{\termA} = \begin{cases}
%%      \COEFvar & \mbox{if $\termA$ is a variable}
%%   \\ \COEFconst & \mbox{if $\termA$ is a constant}
%%   \\ \COEFfunc & \mbox{if $\termA$ is functional}
%%   \\ \COEFpred & \mbox{if $\termA$ is a predicate}
%%   \end{cases}
%%\]
%%%
%%\[
%%   \weightone{\termA} = \begin{cases}
%%   %   \COEFvar & \mbox{if $\termA$ is a variable}
%%   %\\
%%      \COEFconj*\weightzero{\termA} & \mbox{if } \termA\in\related
%%   \\ \weightzero{\termA} & \mbox{otherwise}
%%   \end{cases}
%%\]
%%
%%\heurcode{ConjectureRelativeTermWeight}{$\COEFconj$,$\COEFfunc$,$\COEFconst$,$\COEFpred$,$\COEFvar$,}

\subsubsection{Conjecture Frequency Weight (\Tfidf)}

\emph{Term frequency -- inverse document frequency}, is a numerical
statistic intended to reflect how important a word is to a document
in a corpus~\cite{DBLP:books/cu/LeskovecRU14}.
A \emph{term frequency} is the number of occurrences of the term in a given
document.
A \emph{document frequency} is the number of documents in a corpus
which contain the term.
The term frequency is typically multiplied by the logarithm of the inverse of
document frequency to reduce frequency of terms which appear often.
We define $\tf{\termA}$ as 
the number of occurrences of $\termA$ in $\related$.
We consider a fixed set of clauses denoted $\documents$.
We define $\df{\termA}$ as the count of clauses from $\documents$ which
contain $\termA$.
Out weight function \Tfidf takes one specific argument 
$\COEFdoc$ to select documents, either (1) $\coef{ax}$ for the
axioms or (2) $\coef{pro}$ for all the processed clauses, and
$\WEIGHTONE_\Tfidf$ is as follows.
\small
\[
   \weightbase{\Tfidf} = \frac{1}{1+\tfidf{\termA}}
   \quad\mbox{where}\quad
   \tfidf{\termA} = \tf{\termA}*\log{\frac{1+|\documents|}{1+\df{\termA}}}
\]
\normalsize

%\[\begin{array}{rcl}
%   \tf{\termA} &=& \mbox{``number of occurrences of $\termA$ in $\related$''}
%\\[1mm]
%   \df{\termA} &=& \mbox{``number of $\documents$ which contain $\termA$''}
%\end{array}\]

\subsubsection{Conjecture Term Prefix Weight (\Pref)}

The above weight functions rely on an exact match of a term with a conjecture
related term.
The following weight function loosen this restriction and consider also
partial matches.
We consider terms as symbol sequences.
Let $\pref{\termA}$ be the longest prefix $\termA$ shares with a term
from $\related$.
A \emph{term prefix weight} (\Pref) counts the length of $\pref{\termA}$
using weight arguments $\COEFmatch$ and $\COEFmiss$, formally, 
   $\weightbase{\Pref} = \COEFmatch*|\pref{\termA}| +
   \COEFmiss*(|\termA|-|\pref{\termA}|)$.

%\[
%   \pref{\termA} = \mbox{``the longest prefix $\termA$ shares with a term
%   from $\related$''}
%\]

%\[
%   \weightone{\termA} = \COEFmatch*|\pref{\termA}| + \COEFmiss*(|\termA|-|\pref{\termA}|)
%\]

%\heurcode{ConjectureTermPrefixWeight}{$\COEFmatch$,$\COEFmiss$,}

\subsubsection{Conjecture Levenshtein Distance Weight (\Lev)}

A straightforward extension of \Pref is to employ the Levenshtein
distance~\cite{levenshtein1966bcc} which measures a distance of two strings
as the minimum number of edit operations (character insertion, deletion,
or change) required to change one word into the other.
Our weight function \Lev defines $\weightbase{\Lev}$ as the minimal distance
from $\termA$ to some $\termB\in\related$.
It takes additional arguments $\COEFins$, $\COEFdel$, $\COEFch$ to assign
different costs for edit operations.

%%\[
%%   \lev{\termA}{\termB} = \mbox{``the Levenshtein string distance from $\termA$ to $\termB$''}
%%\]
%%
%%\[
%%   \weightone{\termA} = \min_{\termB\in\related}{\lev{\termA}{\termB}}
%%\]
%%
%\heurcode{ConjectureLevDistanceWeight}{$\COEFins$,$\COEFdel$,$\COEFch$,}

\subsubsection{Conjecture Tree Distance Weight (\Ted)}

The Levenshtein distance does not respect a tree structure of terms.  
To achieve that, we implement the \emph{Tree edit distance} 
\cite{Zhang:1989:SFA:76071.76082} which is similar to Levenshtein but
uses tree editing operations (inserting a node into a tree, deleting a node
while reconnecting its child nodes to the deleted position, and renaming a
node label).
Our weight function \Ted takes the same arguments
as \Lev above and $\WEIGHTONE_\Ted$ is defined similarly.

%\[
%   \ted{\termA}{\termB} = \mbox{``the tree edit distance from $\termA$ to $\termB$''}
%\]
%
%\[
%   \weightone{\termA} = \min_{\termB\in\related}{\ted{\termA}{\termB}}
%\]
%
%\heurcode{ConjectureTedDistanceWeight}{$\COEFins$,$\COEFdel$,$\COEFch$,}

\subsubsection{Conjecture Structural Distance Weight (\Struc)}

With \Ted, a tree produced by the edit operations does not need to represent
a valid term as the operations can change number of child nodes.
To avoid this we define a simple \emph{structural distance} which
measures a distance of two terms by a number of \emph{generalization} and
\emph{instantiation} operations.
Generalization transforms an arbitrary term to a variable while
instantiation does the reverse.
Our weight function \Struc takes additional
arguments $\COEFmiss$, $\COEFinst$, and $\COEFgen$ as penalties for
variable mismatch and operation costs.
The distance of a variable $x$ to a term $\termA$ is the cost of instantiating
$x$ to $\termA$, computed as 
      $\struc{x}{\termA} = \COEFinst*|\termA|$.
The distance of $\termA$ to $x$ is defined similarly but with $\COEFgen$.
A distance of non-variable terms $\termA$ and $\termB$ which share the
top-level symbol is the sum of distances of the corresponding arguments.
Otherwise, a generic formula 
$\struc{\termA}{x_0}+\struc{x_0}{\termB}$
is used.
Function $\WEIGHTONE_\Struc$ is as for \Lev but using $\DSTRUC$.

%
%A distance of a term from/to a variable is defined as follows.
%%
%\small
%\[\begin{array}{lr}
%   \begin{array}{l}
%      \struc{x}{y} = \begin{cases}
%         0 & \mbox{iff }x=y
%      \\ \COEFmiss & \mbox{otherwise}\qquad
%      \end{cases}
%   \end{array}
%&
%   \begin{array}{l}
%      \struc{x}{\termA} = \COEFinst*|\termA|
%   \\ \struc{\termA}{x} = \COEFgen*|\termA|
%   \end{array}
%\end{array}\]
%\normalsize
%%
%A distance of non-variable terms $\termA$ and $\termB$ which share the
%top-level symbol is the sum of distances of corresponding arguments.
%Otherwise, a generic formula 
%$\COEFgen*|\termA|+\COEFinst*|\termB|$
%is used.

%%\[
%%   \struc{f(\termA_1,\dots,\termA_n)}
%%         {f(\termB_1,\dots,\termB_n)} =
%%         \sum_{i=1}^{n}{\struc{\termA_i}{\termB_i}}
%%\]

%\[
%   \weightone{\termA} = \min_{\termB\in\related}{\struc{\termA}{\termB}}
%\]
%
%\heurcode{ConjectureStrucDistanceWeight}{$\COEFmiss$,$\COEFinst$,$\COEFgen$,}

\section{Experimental Results and Evaluation}

The best evaluation would be to measure how our weight functions enrich the
autoschedule mode of E.
This is, however, beyond the scope of this paper.
Instead, we design experiments to help us estimate the quality of the new
weights. 
For each new weight function we run all possible combinations of
common arguments (``$\vartt{v-r-e}$'', see Section~\ref{sec:weights}) and
other manually selected arguments.
First, we run the weight functions on the 2078 MPTP bushy
problems~\cite{abs-1108-3446} with a 5 second time limit.  We compare the
number of solved problems with the number of problems solved by the
\emph{conjecture symbol weight} (denoted $\Ref$) discussed in
Section~\ref{sec:anl}.
Second, to estimate how complementary our weight functions are with existing
functions, we pick a well-performing expert heuristic from the autoschedule
mode of E, and we compute how many problems were solved which the expert
heuristic was not able to solve in 10 seconds (denoted $\Exp$).
The five best-performing combinations of arguments for each weight function are
presented in Table~\ref{tab:experiments}.
Column \emph{speed} contains an average number of processed
\emph{(kilo-)clauses per second} to evaluate implementation efficiency.
Our implementation is available for download\footnote{%
   \texttt{http://people.ciirc.cvut.cz/jakubja5/src/E-arg-2016-03.tar.gz}}.

\begin{table}[p]
\caption{The five best-performing configurations for each weight function.}
\label{tab:experiments}
\begin{center}
\begin{tabular}{l|c|c|c}
\Term & \textbf{solved} & speed & \%\Ref+ \\
\hline                                &              &      & \\[-2ex]
    \variant{\Uni}{\Gen}{\Sum} & \textbf{749} &  5.6 &   5.3 \\
    \variant{\Alf}{\Gen}{\Sum} & \textbf{749} &  5.4 &   5.3 \\
    \variant{\Uni}{\Sub}{\Sum} & \textbf{718} &  5.7 &   1.0 \\
    \variant{\Uni}{\Ter}{\Sum} & \textbf{717} &  5.7 &   0.8 \\
    \variant{\Alf}{\Ter}{\Sum} & \textbf{717} &  5.5 &   0.8 \\\hline
                          \Ref & \textbf{711} &  3.4 &   0.0 \\
\hline
\end{tabular}
\hspace{.5em}
\begin{tabular}{l|c|c|c}
\Term & \textbf{\Exp} & speed & \%\Ref+ \\
\hline                                &              &      & \\[-2ex]
    \variant{\Alf}{\Gen}{\Sim} & \textbf{20} &  4.4 &  -0.7 \\
    \variant{\Uni}{\Sub}{\Sum} & \textbf{19} &  5.7 &   1.0 \\
    \variant{\Uni}{\Ter}{\Sum} & \textbf{19} &  5.7 &   0.8 \\
    \variant{\Alf}{\Ter}{\Sum} & \textbf{18} &  5.5 &   0.8 \\
    \variant{\Alf}{\Sub}{\Sum} & \textbf{18} &  5.5 &   0.6 \\\hline
                          \Ref & \textbf{ 7} &  3.4 &   0.0 \\
\hline
\end{tabular}
\end{center}
\begin{center}
\begin{tabular}{l|c|c|c|c}
\Tfidf & $\COEFdoc$ & \textbf{solved} & speed & \%\Ref+ \\
\hline                         &      &              &      &       \\[-2ex]
    \variant{\Alf}{\Gen}{\Sum} & \coef{ pro} & \textbf{738} &  3.1 &   3.8 \\
    \variant{\Alf}{\Gen}{\Sum} & \coef{  ax} & \textbf{736} &  3.7 &   3.5 \\
    \variant{\Uni}{\Gen}{\Sum} & \coef{ pro} & \textbf{735} &  3.3 &   3.4 \\
    \variant{\Uni}{\Gen}{\Sum} & \coef{  ax} & \textbf{733} &  3.6 &   3.1 \\
    \variant{\Uni}{\Ter}{\Sum} & \coef{ pro} & \textbf{716} &  3.6 &   0.7 \\
\hline
\end{tabular}
\hspace{.5em}
\begin{tabular}{l|c|c|c|c}
\Tfidf & $\COEFdoc$ & \textbf{\Exp} & speed & \%\Ref+ \\
\hline                         &      &              &      &       \\[-2ex]
    \variant{\Uni}{\Sub}{\Sum} & \coef{ pro} & \textbf{17} &  3.5 &   0.3 \\
    \variant{\Uni}{\Gen}{\Sum} & \coef{ pro} & \textbf{16} &  3.3 &   3.4 \\
    \variant{\Uni}{\Ter}{\Sum} & \coef{ pro} & \textbf{16} &  3.6 &   0.7 \\
    \variant{\Alf}{\Sub}{\Sum} & \coef{ pro} & \textbf{16} &  3.3 &   0.1 \\
    \variant{\Uni}{\Sub}{\Sum} & \coef{  ax} & \textbf{16} &  3.9 &   0.0 \\
\hline
\end{tabular}
\end{center}
\begin{center}
\begin{tabular}{l|c|c|c}
\Pref & \textbf{solved} & speed & \%\Ref+ \\
\hline                                &              &      & \\[-2ex]
    \variant{\Alf}{\Gen}{\Sum} & \textbf{788} &  4.0 &  10.8 \\
    \variant{\Alf}{\Top}{\Sum} & \textbf{772} &  4.2 &   8.6 \\
    \variant{\Uni}{\Gen}{\Sum} & \textbf{771} &  3.9 &   8.4 \\
    \variant{\Alf}{\Gen}{\Sim} & \textbf{768} &  3.9 &   8.0 \\
    \variant{\Uni}{\Sub}{\Sum} & \textbf{766} &  4.3 &   7.7 \\
\hline
\end{tabular}
\hspace{.5em}
\begin{tabular}{l|c|c|c}
\Pref & \textbf{\Exp} & speed & \%\Ref+ \\
\hline                                &              &      & \\[-2ex]
    \variant{\Alf}{\Gen}{\Sum} & \textbf{21} &  4.0 &  10.8 \\
    \variant{\Uni}{\Gen}{\Sum} & \textbf{20} &  3.9 &   8.4 \\
    \variant{\Alf}{\Gen}{\Sim} & \textbf{18} &  3.9 &   8.0 \\
    \variant{\Uni}{\Sub}{\Sum} & \textbf{18} &  4.3 &   7.7 \\
    \variant{\Alf}{\Sub}{\Sum} & \textbf{18} &  4.2 &   7.5 \\
\hline
\end{tabular}
\end{center}
\begin{center}
\begin{tabular}{l|c|c|c|c}
\Lev & $\COEF$ & \textbf{solved} & speed & \%\Ref+ \\
\hline                         &      &              &      &       \\[-2ex]
    \variant{\Uni}{\Gen}{\Sim} & \coef{ 155} & \textbf{841} &  2.4 &  18.3 \\
    \variant{\Alf}{\Gen}{\Sim} & \coef{ 155} & \textbf{836} &  2.4 &  17.6 \\
    \variant{\Alf}{\Gen}{\Sim} & \coef{ 151} & \textbf{827} &  2.5 &  16.3 \\
    \variant{\Alf}{\Gen}{\Sim} & \coef{ 111} & \textbf{824} &  2.5 &  15.9 \\
    \variant{\Uni}{\Gen}{\Sim} & \coef{ 151} & \textbf{822} &  2.5 &  15.6 \\
\hline
\end{tabular}
\hspace{.5em}
\begin{tabular}{l|c|c|c|c}
\Lev & $\COEF$ & \textbf{\Exp} & speed & \%\Ref+ \\
\hline                         &      &              &      &       \\[-2ex]
    \variant{\Uni}{\Gen}{\Sim} & \coef{ 155} & \textbf{41} &  2.4 &  18.3 \\
    \variant{\Alf}{\Gen}{\Sim} & \coef{ 155} & \textbf{39} &  2.4 &  17.6 \\
    \variant{\Alf}{\Gen}{\Sim} & \coef{ 151} & \textbf{35} &  2.5 &  16.3 \\
    \variant{\Alf}{\Gen}{\Sim} & \coef{ 111} & \textbf{35} &  2.5 &  15.9 \\
    \variant{\Uni}{\Gen}{\Sim} & \coef{ 151} & \textbf{30} &  2.5 &  15.6 \\
\hline
\end{tabular}
\end{center}
\begin{center}
\begin{tabular}{l|c|c|c|c}
\Ted & $\COEF$ & \textbf{solved} & speed & \%\Ref+ \\
\hline                         &      &              &      &       \\[-2ex]
    \variant{\Alf}{\Gen}{\Sim} & \coef{ 511} & \textbf{797} &  1.2 &  12.1 \\
    \variant{\Alf}{\Gen}{\Sim} & \coef{ 111} & \textbf{797} &  1.3 &  12.1 \\
    \variant{\Uni}{\Gen}{\Sum} & \coef{ 155} & \textbf{789} &  1.0 &  11.0 \\
    \variant{\Alf}{\Gen}{\Sum} & \coef{ 155} & \textbf{789} &  1.0 &  11.0 \\
    \variant{\Alf}{\Gen}{\Sim} & \coef{ 155} & \textbf{789} &  1.2 &  11.0 \\
\hline
\end{tabular}
\hspace{.5em}
\begin{tabular}{l|c|c|c|c}
\Ted & $\COEF$ & \textbf{\Exp} & speed & \%\Ref+ \\
\hline                         &      &              &      &       \\[-2ex]
    \variant{\Alf}{\Gen}{\Sim} & \coef{ 111} & \textbf{33} &  1.3 &  12.1 \\
    \variant{\Alf}{\Gen}{\Sim} & \coef{ 511} & \textbf{32} &  1.2 &  12.1 \\
    \variant{\Alf}{\Gen}{\Sim} & \coef{ 155} & \textbf{28} &  1.2 &  11.0 \\
    \variant{\Uni}{\Gen}{\Sum} & \coef{ 155} & \textbf{25} &  1.0 &  11.0 \\
    \variant{\Uni}{\Ter}{\Sim} & \coef{ 511} & \textbf{23} &  2.4 &   6.2 \\
\hline
\end{tabular}
\end{center}
\begin{center}
\begin{tabular}{l|c|c|c|c}
\Struc & $\COEF$ & \textbf{solved} & speed & \%\Ref+ \\
\hline                         &      &              &      &       \\[-2ex]
    \variant{\Uni}{\Ter}{\Sim} & \coef{ 115} & \textbf{833} &  3.9 &  17.2 \\
    \variant{\Alf}{\Ter}{\Sim} & \coef{ 115} & \textbf{832} &  2.0 &  17.0 \\
    \variant{\Uni}{\Sub}{\Sum} & \coef{ 115} & \textbf{832} &  2.9 &  17.0 \\
    \variant{\Alf}{\Sub}{\Sum} & \coef{ 115} & \textbf{831} &  1.4 &  16.9 \\
    \variant{\Uni}{\Sub}{\Sim} & \coef{ 115} & \textbf{825} &  3.6 &  16.0 \\
\hline
\end{tabular}
\hspace{.5em}
\begin{tabular}{l|c|c|c|c}
\Struc & $\COEF$ & \textbf{\Exp} & speed & \%\Ref+ \\
\hline                         &      &              &      &       \\[-2ex]
    \variant{\Uni}{\Sub}{\Sum} & \coef{ 115} & \textbf{32} &  2.9 &  17.0 \\
    \variant{\Alf}{\Sub}{\Sum} & \coef{ 115} & \textbf{32} &  1.4 &  16.9 \\
    \variant{\Alf}{\Top}{\Sum} & \coef{ 115} & \textbf{31} &  1.5 &  16.0 \\
    \variant{\Uni}{\Ter}{\Sim} & \coef{ 115} & \textbf{29} &  3.9 &  17.2 \\
    \variant{\Uni}{\Top}{\Sum} & \coef{ 115} & \textbf{29} &  2.9 &  15.6 \\
\hline
\end{tabular}
\end{center}
\end{table}

%and Table~\ref{tab:experiments} provides experimental evaluation featuring five 
%manually found best combinations of arguments 
%   $\vartt{v}\in\{\Alf,\Uni\}$, 
%   $\vartt{r}\in\{\Ter,\Sub,\Top,\Gen\}$, and
%   $\vartt{e}\in\{\Sim,\Sum,\Max\}$ (written as ``$\vartt{v-r-e}$'').
%%All the experiments were run with 
%%    $\COEFmaxterm=\COEFmaxlit=\COEFposmult=1.5$ and with a constant priority
%%    function \procStyle{ConstPrio} which technically disables the priority
%%    function.
%The performance is measured by the number of the 2078 MPTP bushy 
%problems~\cite{abs-1108-3446} solved within a 5 second time limit.
%Additionally, we consider how many problems were solved which the expert
%heuristic TODO was not able to solve in 10 seconds
%time limit (column $\Exp$).
%This gives us a measure of diversity of the heuristics. % solved problems.
%For each run we compute a percentual gain in terms of the number of solved problems over the reference
%heuristic $\Ref$ (column $\%\Ref+$).

From Table~\ref{tab:experiments} we can see that the weights which rely
on an exact match of a term with a related term or its part (\Term, \Tfidf,
and \Pref) perform best when values of $\WEIGHTONE$ are summed for all the
subterms ($\vartt{e}=\Sum$).
On the other hand, weights which incorporate some notion of term
similarity directly in $\WEIGHTONE$ do not profit so much from this.
%%The experiments further reveal that the variants with unified variables
%($\vartt{v}=\Uni$) perform in many cases equally to the corresponding
%variants with $\alpha$-normalized variables.
%As for the variations on the set $\related$, in many cases all subterms
%generalizations ($\Gen$) perform best, however, the success of other variants
%suggests that they are also worth of considering.
For weights \Lev, \Ted, and \Struc we have tried to experiment with
operation costs (column $\COEF$, for example, $\coef{151}$ means that
$\COEFdel$ is increased to $5$ while other costs are $1$).
In general, the experiments show that different arguments have an impact 
on performance.
% Finding the best heuristic arguments is left for the future research where
% we plan to employ automated machine learning methods~\cite{DBLP:journals/corr/abs-1301-2683}.
Finally, the experiments also reveal a higher time complexity of the \Lev and \Ted
weights (Levenshtein distance of two terms is in $O(n^2)$ while
\Ted is in $O(n^3)$).
However, a higher time complexity does not have to be a drawback as \Lev is
still best performing.

\section{Conclusions and Future Work}

We have implemented several new weight functions for E prover based on term
similarity with a conjecture.
The experiments suggest that our functions have a potential to improve the
autoschedule mode of E as they are reasonably complementary with existing
heuristics. 
In order to use our weight functions with the autoschedule mode of E, we
would need to (1) find the best performing parameters of our weight
functions, (2) find the best combinations of our weight functions with other
weight functions, and (3) find the most complementary combinations and
create a scheduling strategy.  
As a future research, we are planning to use
parameter-searching methods such as
BliStr~\cite{GCAI2015:BliStr_The_Blind_Strategymaker} to achieve this task.

\bibliographystyle{plain}
\bibliography{eheuristics-CICM16}

\end{document}